\documentclass[
aps,
prl,
twocolumn,
letterpaper,
superscriptaddress,
10pt
]{revtex4-2}

\usepackage{amssymb,amsthm,amsmath,amsfonts}
\usepackage{graphicx,enumerate,bbm,bm,mathptmx}
\usepackage[pdftex,dvipsnames,usenames]{xcolor}
\usepackage{soul}
\usepackage[colorlinks=true,urlcolor=blue,citecolor=blue,linkcolor=blue]{hyperref}

\begin{document}

% \title{Universal correlations}
% universal correlations: classical versus quantum
% Entanglement-Inspired Universal Correlations
% Unified Framework for Entanglement-Based Correlations
\title{Universal entanglement-inspired correlations}
% further suggestions???

\author{Elizabeth Agudelo}
\email{elizabeth.agudelo@tuwien.ac.at}
    \affiliation{TU Wien, Atominstitut \& Vienna Center for Quantum Science and Technology, Stadionallee 2, 1020 Vienna, Austria}

\author{Laura Ares}
\email{laura.ares.santos@uni-paderborn.de}
    \affiliation{Paderborn University, Institute for Photonic Quantum Systems (PhoQS), Theoretical Quantum Science, Warburger Stra\ss{}e 100, 33098 Paderborn, Germany}

\author{Jan Sperling}
\email{jan.sperling@uni-paderborn.de}
    \affiliation{Paderborn University, Institute for Photonic Quantum Systems (PhoQS), Theoretical Quantum Science, Warburger Stra\ss{}e 100, 33098 Paderborn, Germany}
\date{\today}

\begin{abstract}
    Quantum correlations, crucial for the advantage and advancement of quantum science and technology, arise from the impossibility of expressing a quantum state as a tensor product over a given set of parties.
    In this work, a generalized notion of correlations via arbitrary products is formulated.
    Remarkably, as a universal property, the connection between such general products and tensor products is established, allowing one to relate generic non-product states to the common notion of entangled states.
    We construct the set of free operations for general types of products by extending the local-operation-and-classical-communication paradigm, familiar from standard entanglement theory, thereby establishing a resource theory of correlations for general products.
    A generalization is provided beyond two factors that can be universally related to multipartite entanglement.
    Applications that highlight the usefulness of the approach are discussed, such as the factorization of fermionic states, the non-local factorization of multi-photon states into single-photon states, and the interesting possibility of understanding prime numbers as a form of single-party entanglement.
\end{abstract}

\maketitle

\paragraph{Introduction.---}
    Quantum correlations, particularly entanglement, are the defining quantum features that fuel not only our modern understanding of nature, but also the development of modern quantum technologies \cite{GT09,HHHH09,FVMH19}. 
    They enable capabilities far beyond what is achievable with classical physical systems \cite{B64,CHSH69,BCPSW14}. 
    The existence of entanglement has long been demonstrated in pioneering experiments \cite{AGR81,ADR82} and, more recently, in loophole-free implementations \cite{Hetal15,Getal16,Setal16}, including the development of device-independent verification techniques \cite{BRLG13}. 
    The benefits offered by entanglement have been established in quantum metrology \cite{T12,HLKSWWPS12,M13,OAGKAL16}, quantum computation and communication \cite{NC00,WPGCRSL12}.
    It has also been observed that the structure of the entanglement between parties can be more relevant than its amount for certain protocols \cite{DFC05,GFE09}.
    Moreover, ongoing efforts aim to uncover its connections to other areas of physics \cite{H16,C15}, resulting in applications beyond the purely academic sphere \cite{DM03}.
    
    The relationship between entanglement and other indicators of quantumness has been well established \cite{AFLP66,KSBK02,X02,WEP03,VS14} and multiple unification approaches have been proposed \cite{MPSVW10,AACHKLMP10,TKPJ16,GZ18,CPV19,KASSVSH21}. 
    Beyond entanglement, other forms of quantum correlation have also been shown to play essential roles in quantum information processing and quantum optics, where they are crucial in distinguishing the classical or quantum nature of correlations \cite{DSC08,FP12,ASV13,MBCPV12,SAWV17}. 
    Entanglement theory has by now evolved into a field of its own, and the ongoing exploration of its far-reaching impact across different areas of physics only reinforces the central role of this quantum phenomenon \cite{BengtssonZyczkowski_2006, NC00, BuchleitnerViviescasTiersch_2009, FurusawaVanLoock_2011, Brody_2020}.
    
    Entanglement arises in composite systems that contain at least two subsystems which cannot be described independently---defining the notion of non-separable \cite{W89}. 
    Although its definition is simple, entanglement is remarkably challenging to characterize and can only be identified when the subsystems involved are analyzed jointly \cite{AFOV08,HV13,LM13,SSV14,GSVCRTF16}. 
    Specifically, a product state is a composite quantum state that can be written as a tensor product of states of its subsystems.
    A quantum state is separable if it can be expressed as a product state (e.g., for pure states) or a mixture of product states (for mixed states), allowing it to be prepared using only local operations and classical communication (LOCC) \cite{N99,CLMOW14}. 
    In contrast, entangled states cannot be generated by LOCC.
    Determining whether a general mixed state is entangled is challenging because it requires ruling out all possible decompositions into mixtures of product states, a problem that is NP-hard even under relaxed error tolerances \cite{G03,G04,G10}.

    Today, entanglement is often studied within the framework of quantum resource theories \cite{CG19}, connecting it to broader notions, such as quantum coherence, and highlighting its operational usefulness in quantum protocols. 
    In this context, surpassing free operations---LOCC in the case of entanglement---demonstrates a system’s ability to outperform any classical counterpart. 
    There are manifold of ways in which this quantum property of non-separability can manifest itself \cite{SA23,SPBS19,SW17,BASS25,SGBA23,PDBBSS21}.
    Certifying entanglement is therefore a complex and multifaceted challenge, and the techniques developed for its analysis may offer insights with applications beyond physics \cite{GT09}. 
    Whether such diversity extends to other forms of correlations---non-factorizability in terms of different products---remains an open question, one that we address in this work through establishing a unified framework.
    
    In this paper, we investigate the ``quantum-factorization'' problem for generic types of products. 
    One of our central results is that all non-factorizable states, with respect to a given product, can be connected to the notion of quantum entanglement through a unique linear map. 
    We extend this result to an arbitrary number of factors and to mixed states, by allowing classical correlations within ensembles of multi-factor product states. 
    Furthermore, the universal property we establish allows us to generalize the LOCC paradigm from standard entanglement theory to arbitrary products of states, thereby defining the corresponding resource-free operations for the product under study.    
    Finally, we illustrate the applicability of our unified framework through a range of examples, including quantum correlations in fermionic and bosonic systems, correlations within a single degree of freedom, and even a reformulation of the problem of finding prime numbers as an entanglement verification problem.

\paragraph{General product states.---}

	Let $|\psi_A\rangle$ and $|\psi_B\rangle$ be two quantum states of the systems $A$ and $B$, respectively.
    A general product $\circ$ is given by the map
    \begin{equation}
        \circ:(|\psi_A\rangle,|\psi_B\rangle)\mapsto |\psi_A\circ\psi_B\rangle
    \end{equation}
    that is bilinear---but not necessarily the tensor product---that outputs the product vector $|\psi_A\circ\psi_B\rangle$.
	Let us recall that a bilinear operation obeys the left and right distributive property as well as associativity with respect to the multiplication with a scalar
    \footnote{Bilinearity consists of the additive/distributive property, $|(a+a')\circ b\rangle=|a\circ b\rangle+|a'\circ b\rangle$ and $|a\circ(b+b')\rangle=|a\circ b\rangle+|a\circ b'\rangle$, and the homogeneous/distributive property, $\lambda|a\circ b\rangle=|(\lambda a)\circ b\rangle=|a\circ(\lambda b)\rangle$ for all scalars $\lambda$.}.
	If a vector $|\psi_{AB}\rangle$ can be written as $|\psi_A\circ\psi_B\rangle$, then, by definition, we refer to $|\psi_{AB}\rangle$ as $\circ$-factorizable;
	otherwise, we call $|\psi_{AB}\rangle$ $\circ$-entangled, i.e.,
	\begin{equation}
	\begin{aligned}
		|\psi_{A|B}\rangle=|\psi_A\circ\psi_B\rangle
		\quad&\Leftrightarrow\quad\text{$\circ$-factorizable},
		\\
		|\psi_{AB}\rangle\neq|\psi_A\circ\psi_B\rangle
		\quad&\Leftrightarrow\quad\text{$\circ$-entangled}.
	\end{aligned}
	\end{equation}
	This introduces a notion of entanglement that is not limited to tensor products and allows us to speak about correlations for all types of bilinear products;
	multilinear generalizations are discussed below.
	Although our generalized definition of products is not restricted to tensor products, we show next that it can be related to the common notion of entanglement via the following interesting observation.

\paragraph{Universality.---}

\begin{figure}[t]
    \centering
    \includegraphics[width=0.9\columnwidth]{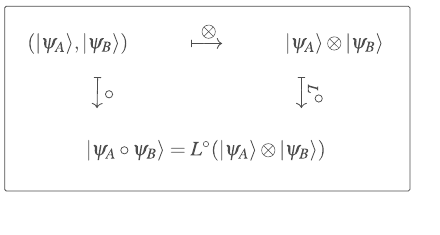}
    \caption{%
        Universal property as commuting diagram.
        The factors combined with the bilinear product $\circ$ result in the same outcome as the linear map $L^\circ$ acting on their tensor product $\otimes$.
    }\label{fig:universal}
\end{figure}

	The so-called universal property of the tensor product $\otimes$ is that there exist a unique, linear map acting on the tensor-product space for each bilinear map that acts on the individual subspaces \cite{R07};
    cf. Fig. \ref{fig:universal}.
	Recalling that the product $\circ$ is a bilinear map, we thus can formulate the following, universally true observation:
	there exists a unique, linear operator $L^\circ$ such that
	\begin{equation}
		|\psi_A\circ\psi_B\rangle=L^\circ(|\psi_A\rangle\otimes|\psi_B\rangle)
	\end{equation}
	holds true for all $\circ$-factorizable states.
	This universality enables us to relate the common notion of entangled states in terms of $\otimes$ to the broader concept of $\circ$-entanglement of a state $|\psi_{AB}\rangle$, i.e., $|\psi_{AB}\rangle\neq L^\circ(|\psi_A\rangle\otimes|\psi_B\rangle)$.

	As a consequence of universality, all resources of quantum correlations as broadly defined above can be represented---up to the map $L^\circ$---as a consequence of $\otimes$-entanglement.
	This has significant impact on the theoretical and experimental verification and quantification of $\circ$-entanglement which now can be based on the existing approaches from $\otimes$-entanglement theory by modifying them through the map $L^\circ$.
	Also note the simple fact that standard $\otimes$-entanglement theory is the special case in which $L^\otimes=\mathbbm 1$ (the identity map) holds.

\paragraph{Paradigm of factor-wise transformations.---}

	Next, suppose we have transformations $|\psi_S\rangle\mapsto T_S|\psi_S\rangle=|T_S\psi_S\rangle$ for $S\in\{A,B\}$.
    Considering universality, we find the preservation of $\circ$-factorization as follows:
	\begin{equation}
		L^\circ\left(
			\left[T_A\otimes T_B\right]
			\left[|\psi_A\rangle\otimes|\psi_B\rangle\right]
		\right)
		=
		|(T_A\psi_A)\circ(T_B\psi_B)\rangle.
	\end{equation}
	In the language of $\otimes$-entanglement, we know that local transformations, such as $T_A\otimes T_B$, are free operations, which play a crucial role for general resource theories.
	The above identity proves that we find the analogous paradigm for $\circ$-factorization as it is {closed} under factor-wise transformations, $L^\circ\, (T_A\otimes T_B)$, thus not producing a quantum resource.

\paragraph{Convex hull.---}

	Through statistical ensembles, we can extend the notion of $\circ$-factorizable states to define classically $\circ$-correlated states via the {closure} of the convex hull of pure states. 
	That is, a classically $\circ$-correlated state takes the form
	\begin{equation}
		\sigma_{A|B}=\int dP(\psi_A,\psi_B)\frac{|\psi_A\circ\psi_B\rangle\langle \psi_A\circ\psi_B|}{\langle \psi_A\circ\psi_B|\psi_A\circ\psi_B\rangle},
	\end{equation}
	where $P$ is a joint probability distribution and the denominator is included to ensure normalization of the pure product states.
    With the notation $A|B$, we denote quantum-factorizability.
	If a state $\varrho_{AB}$ is not in the closure of the convex hull, it is defined to be quantum $\circ$-correlated.

	Note that an affine resource theory can be constructed when allowing for signed measures $P$---i.e., including negative quasiprobabilities---as long as $\sigma_{A|B}$ remains a valid quantum state.
	Because of the universality property mentioned above, classical $\circ$-correlations directly relate to the notion of a $\otimes$-separable state $\sigma_\mathrm{sep}$ \cite{W89}, i.e., $\sigma_{A|B}=L^\circ \sigma_\mathrm{sep} L^{\circ\dag}$.
	In terms of factor-wise transformations and universality, the LOCC paradigm of inseparability now directly maps to the corresponding paradigm for free operations on mixed, classically $\circ$-correlated states.

\paragraph{Multi-factor correlations.---}

	In addition to the product of two factors, one can extend the concept of a $\circ$-factorizable states to multilinear maps, e.g., $|\psi_A\circ\psi_B\circ\psi_C\rangle$ for three factors.
	Again, the universality of the tensor product renders it possible to find a linear map $L^\circ$ such that $|\psi_A\circ\psi_B\circ\psi_C\rangle=L^\circ(|\psi_A\rangle\otimes|\psi_B\rangle\otimes|\psi_C\rangle)$ holds true, further connecting the commonly applied notion of multipartite entanglement and the here-introduced concept of $\circ$-entanglement with respect to multiple factors.
	Using the factorization with as many factors as needed, one can also think about partial $\circ$-entanglement, e.g., $|\psi_{AB}\circ\psi_B\rangle=L^\circ(|\psi_{AB}\rangle\otimes|\psi_C\rangle)$ for $|\psi_{AB}\rangle\neq |\psi_A\circ\psi_B\rangle$.
	Introducing classical correlations via mixing, if a state $\varrho_{ABC}$ fails to be a $\circ$-biseparable state, $p_{AB|C}\sigma_{AB|C}+p_{BC|A}\sigma_{BC|A}+p_{CA|B}\sigma_{CA|B}$ where $p_{AB|C},p_{BC|A},p_{CA|B}\geq0$ and $p_{AB|C}+p_{BC|A}+p_{CA|B}=1$, the state $\varrho_{ABC}$ exhibits the $\circ$-analog to genuine multipartite entanglement \cite{GT09,HHHH09}.

\paragraph{Examples.---}

	In the following, we provide fundamental and more advanced examples for the quantum correlations introduced.
	$\otimes$-entanglement, where $L^\otimes=\mathbbm 1$, is a trivial instance for our approach, thus not further discussed.
	The selection of known and unknown products to which our approach applies demonstrates the versatility of the proposed framework.

	Considering two fermions in first quantization, the vector $|0\wedge 1\rangle/\sqrt2=(|0\rangle\otimes|1\rangle-|1\rangle\otimes |0\rangle)/\sqrt2$, where $\wedge$ denotes the exterior product, can be classified as $\wedge$-factorizable but is, at the same time, an $\otimes$-entangled Bell state.
	Within our framework, the debate whether exchange symmetry counts as a form of entanglement is answered by: it depends on the choice of product that is used to assess quantum correlations.
    For instance, the resourcefulness of superpositions of identical particles was demonstrated in Ref. \cite{MYFZTA20}.

	To provide a non-trivial, single-system, continuous-variable example for quantum correlations with more than two factors, we consider $|\psi_A\rangle=\sum_{n=0}^\infty a_{n}|n\rangle$, $|\psi_B\rangle=\sum_{n=0}^\infty b_{n}|n\rangle$, and $|\psi_C\rangle=\sum_{n=0}^\infty c_{n}|n\rangle$ together with the product $\circ$, defined in terms of the coefficients of the factors of these three states as
	\begin{equation}
		|\psi_{A|B|C}\rangle=\left(\sum_{n=0}^\infty a_{n}b_{n}c_{n}\right)|0\rangle +\sum_{n=0}^\infty a_{n}b_{n}c_{n}|n+1\rangle.
	\end{equation}
	It is straightforward to verify that this constitutes a trilinear product.
	In addition, we take $|\psi_{ABC}\rangle=\sum_{n} q^n |n\rangle$, with $0<q<1$ and ignoring a proper normalization as it does not affect factorization.
    Note that this state is defined over a single party with the same basis $\{|n\rangle:n\in\mathbb N\}$ as used for each factor.
	To probe the $\circ$-correlations, we equate the coefficients of $|\psi_{A|B|C}\rangle$ and $|\psi_{ABC}\rangle$, resulting $q^{n+1}=a_nb_nc_n$ $\forall n$ and $1=\sum_{n} a_n b_nc_n$ for the zeroth component.
	Combining the two identities, we obey $\circ$-factorizability for $q=1/2$ and find $\circ$-entanglement otherwise because $1=\sum_{n=0}^\infty q^{n+1}=q/(1-q)$	implies $q=1/2$ as the only valid solution.
	Thus, we constructed a proof-of-concept example that shows tripartite $\circ$-entanglement for all $q\in(0,1)\setminus\{1/2\}$ within a single, infinite-dimensional system.

	In a bosonic system in second quantization, two single photons spread across $M$ modes can be written as $|\psi_A\rangle=\sum_{m=1}^M\alpha_m\hat a_m^\dag|\mathrm{vac}\rangle$ and $|\psi_B\rangle=\sum_{n=1}^M\beta_n\hat a_n^\dag|\mathrm{vac}\rangle$, where $\hat a^\dag_m$ is the creation operator for the $m$\textsuperscript{th} mode and $|\mathrm{vac}\rangle$ is the vacuum state.
	Exciting both states defines a two-photon product vector, 
	\begin{equation}
		\label{eq:PhotonProduct}
		  |\psi_{A|B}\rangle
		=
		\sum_{m=1}^M \alpha_m\hat a^\dag_{m}
		\sum_{n=1}^M \beta_n\hat a^\dag_{n}
		|\mathrm{vac}\rangle,
	\end{equation} 
    even if the factors are not locally restricted to one specific mode, meaning that the photons factorize, but the states do not necessarily factorize in modes.
    The map $L^\circ$ can be defined through $L^\circ(\hat a_m^\dag|\mathrm{vac}\rangle\otimes \hat a^\dag_n|\mathrm{vac}\rangle)=\hat a^\dag_m\hat a^\dag_n|\mathrm{vac}\rangle$.
	Now suppose $M=4$ modes and $|\psi_{AB}\rangle=(\hat a^\dag_2\hat a^\dag_4+\hat a^\dag_1\hat a^\dag_3)|\mathrm{vac}\rangle=|0,1,0,1\rangle+|1,0,1,0\rangle$.
	Equating the coefficients with a product state shows that the two photons are entangled (for details on the calculation cf. comment
    \footnote{Equating the coefficients in Eq. \eqref{eq:PhotonProduct} for $M=4$ with $(\hat a^\dag_2\hat a^\dag_2+\hat a^\dag_1\hat a^\dag_3)|\mathrm{vac}\rangle$ can be summarized via the terms $T_{m,n}=\alpha_m\beta_n+\alpha_n\beta_m$, where $m,n\in\{1,2,3,4\}$ and $m\leq n$, that ought to be zero for all pairs $(m,n)$ except for $(m,n)\in\{(1,3),(2,4)\}$.
		This defines a system of equations that has no solution, which can be shown as follows:
		$T_{1,1}=0$ $\stackrel{\text{w.l.o.g.}}{\Rightarrow}\alpha_1=0$;
		$T_{1,3}\neq0$ $\Rightarrow$ $\alpha_3\neq0$ \& $\beta_1\neq0$;
		$T_{3,3}=0$ $\Rightarrow$ $\beta_3=0$;
		$T_{1,4}=0$ $\Rightarrow$ $\alpha_4=0$;
		$T_{2,3}=0$ $\Rightarrow$ $\beta_2=0$;
		$T_{2,4}\neq0$ $\Rightarrow$ $\alpha_2\neq0$ \& $\beta_4\neq0$;
		and the equations $T_{1,2}=0$ is in contradiction to the found parameters that yield $T_{1,2}\neq0$, likewise for $T_{3,4}$.
		(The chain of arguments with swapped $\alpha$s and $\beta$s holds true when taking $\beta_1=0$ in the first step.)}%
    ) with respect to the product defined in Eq. \eqref{eq:PhotonProduct}.
	We emphasize that factorization is considered here with respect to the delocalized photon states $|\psi_A\rangle$ and $|\psi_B\rangle$ \cite{KOAWOMSS25}, contrasting factorization with respect to states of local modes, which is the common approach.

\begin{figure}[t]
    \centering
    \includegraphics[width=0.9\columnwidth]{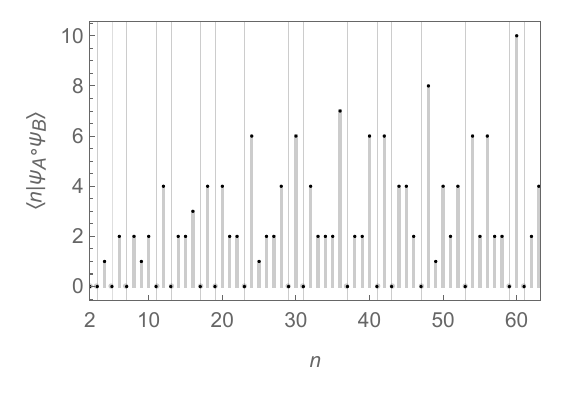}
    \caption{%
        Product state for product $|n_A\rangle\circ|n_B\rangle=|n_A\cdot n_B\rangle$, where we consider unnormalized states $|\psi_A\rangle=|\psi_B\rangle=\sum_{n\geq 2} 1 |n\rangle$ as factors describing Alice's and Bob's state.
        The resulting components of $|\psi_A\circ\psi_B\rangle=\sum_{n\geq 2} c_n |n\rangle$ are shown, where $c_n=\langle n|\psi_A\circ\psi_B\rangle$.
        Vertical lines indicate prime $n$, occurring with a zero amplitudes in product states.
        In general, the amplitude tells us the number of ways one can write as $n=n_A\cdot n_B$, with $n_A,n_B\geq 2$.
    }\label{fig:primes}
\end{figure}

	The final example---motivated purely as a computational application---is similar to the second one since we consider a single system.
    Here, however, we want to study prime numbers in the context of our entanglement-inspired theory of universal correlations.
    With the orthonormal basis $\{|n\rangle: 2 \leq n\in\mathbb N\}$, we can define $|\psi_A\circ\psi_B\rangle=\sum_{m,n=2}^\infty a_m b_n |m\cdot n\rangle$, where $|\psi_A\rangle=\sum_{m=2}^\infty a_m|m\rangle$, $|\psi_B\rangle=\sum_{n=2}^\infty b_n |n\rangle$, and $\cdot$ denoting the normal products between integers.
	We can also expand
	\begin{equation}
		|\psi_{A|B}\rangle
		=\sum_{q=2}^\infty
		%\underbrace{
        \left(
			\sum_{m,n\geq 2:m\cdot n=q} a_m b_n
		\right)%}_{=c_q}
		|q\rangle
        =\sum_{q=2}^\infty
		c_q
		|q\rangle.
	\end{equation}
	Note that the sum $c_q$ in the parentheses is empty, i.e., zero, for $q$ being a prime number.
	Thus, all basis states $|p\rangle$ for prime numbers $p$ are $\circ$-entangled.
	Furthermore, if the coefficients of both factors are constant up to $N$ and zero afterwards, we have $c_q\propto\sum_{m,n\in\{2,\ldots,N\}:m\cdot n=q}1$, being proportional to the possible factorizations of $q$ in terms of two (non-zero and non-unit) integer factors;
    see Fig. \ref{fig:primes}.
    
\paragraph{Conclusion.---}
  
    In this work, we introduced a unified framework that generalizes quantum correlations beyond the tensor-product structure by defining general correlations relative to arbitrary products.
    Within this generalized notion, the analog to non-entangled states admits a factorization into at least two non-trival factors for the product under study.
    Further, a universal correspondence between generalized products and the tensor product establishes a direct link between non-factorizable states and conventional entanglement.
    This connection enables the construction of the associated set of free operations through a natural extension of the LOCC paradigm.
    The corresponding resource theory emerges then naturally from our construction, offering an operational treatment for studying generalized correlations beyond entanglement.

    We extend our notion of factorizability to mixed-state scenarios, defining classically correlated states with respect to arbitrary products.
    Moreover, we show how our framework applies to multifactor products, i.e., multilinear maps, which directly relate to the theory of multipartite entanglement, recovering concepts such as the analogs to partial entanglement and genuine multipartite entanglement.
    Overall, our approach provides a consistent description of generalized correlations for complex systems and generalized notions of factorization.
    
    Compelling applications we discussed include the anti-symmetric product relevant to fermionic systems in first quantization and the nonlocal factorization of photon states in second quantization.
    Furthermore, we anticipate even more intriguing uses of this framework, as illustrated by interpreting prime-number states as a form of single-party entanglement.
    For instance, one can imagine in this regard future construction of useful algorithms, using resources defying generalized products, such as extensions of Shor's algorithm \cite{S97} that is currently limited to prime-number factorization.
    Transcending the standard notion of separability in entanglement theory, our approach renders it possible to map factorization problems onto each other, enabling the powerful tools developed for quantum entanglement theory to be employed in a wide variety of unconventional contexts.
    
\paragraph{Acknowledgements.---}%
    The authors are grateful to Esteban Vasquez for his insightful observations and valuable comments.
    E.A. acknowledges funding from the Austrian Science Fund (FWF) through the Elise Richter project 10.55776/V1037.
    L.A. and J.S. acknowledge funding through the Quant\-ERA project QuCABOoSE.

\bibliographystyle{apsrev-new}
\bibliography{UnivCorrel}% Produces the bibliography via BibTeX.

\end{document}